\documentclass[runningheads]{svmult}

\usepackage{makeidx}   
\usepackage{graphicx}  
\usepackage{subeqnar}  
\usepackage{multicol}  
\usepackage{physmubb}  
\makeindex             

\begin{document}
\title*{Jet Physics at HERA}
\toctitle{Jet Physics at HERA}
%
%
\titlerunning{Jet Physics at HERA}
%
\author{Oscar Gonz\'alez\\(On behalf of the H1 and ZEUS Collaborations)}

\authorrunning{Oscar Gonz\'alez}
%
%
\maketitle              

\vspace{-0.5cm}
\centerline{{\bf Abstract}}
\vspace{0.1cm}
\noindent
Analyses of jet production at HERA have become
precise enough to perform QCD studies which are both 
competitive and complementary to those at $e^+e^-$ and $p\overline{p}$
colliders.
This report summarises 
some of the latest results on jet physics from the H1 and ZEUS
Collaborations. 

\section{Introduction}

The high-energy $ep$ interactions at the HERA collider provide a powerful 
laboratory to test the prediction of 
Quantum Chromo-Dynamics~(QCD). 
In this theory, the interactions between quarks and gluons 
produce partons with large transverse momenta, 
which fragment into hadronic jets.

In recent years the experimental
uncertainties have been substantially 
reduced 
and the tools
to compute the theoretical predictions have been refined so that
more
accurate QCD studies are possible with jet physics at HERA.

This report presents a selection of the latest results obtained
by the H1 and ZEUS Collaborations at the HERA collider in both 
the photoproduction and deep inelastic scattering~(DIS) regimes.

\section{Photoproduction of Jets}

The photoproduction of jets with high transverse energy 
is described by the hard interactions of real photons with the partons
inside the proton. These interactions are separated into two classes: 
direct processes, in which the photon interacts as a point-like particle 
with a parton in the proton, and resolved processes, in which the 
photon has an hadronic structure and one of the partons in the photon
scatters a parton in the proton. In order to separate
these two processes, the variable $x_\gamma$ is used. This quantity is the fraction
of the photon's momentum which takes part in the interaction. It is expected 
that high values of $x_\gamma$, i.e. close to unitiy, are related to direct 
processes, whilst the resolved processes have a 
distribution on this variable at lower $x_\gamma$ values.

\subsection{The Internal Structure of the Photon}

The study of dijet production in the photoproduction regime gives information 
on the partonic content of the photon as well as providing a test of the 
QCD predictions.
The variable $x_\gamma$ is reconstructed 
from the transverse energies and 
pseudorapidities of the two jets with the highest transverse energies:
\begin{equation}
x_\gamma = \frac{1}{2\,E_\gamma} \big[ E_{T,{\rm jet1}} \exp ({-\eta_{\rm jet1}})  
                      +   E_{T,{\rm jet2}} \exp({-\eta_{\rm jet2}}) \big]\;,
\end{equation}
where $E_\gamma$ is the energy of the incoming quasi-real photon.

Measurements of dijet cross sections in the photoproduction regime are 
sensitive to both proton and photon parton distribution functions (pdfs),
as shown in the analyses of the H1 and ZEUS 
Collaborations~\cite{h1dijphp,zeusdijphp}. The data are reasonably well 
described 
by the next-to-leading order (NLO) QCD predictions.
Due to the small uncertainties in the proton pdfs, well constrained by the
DIS data, measurements of dijet photoproduction can be used to study
the structure of the photon.

In order to use these data to precisely constrain the photon pdfs, a reduction of the
theoretical scale uncertainties, possibly via the inclusion of higher orders, is
needed~\cite{h1dijphp,zeusdijphp}. 
At present, the data are only used to test the existing parameterisations.

Apart from testing the photon pdfs, dijet photoproduction has also been
used to perform searches of high-mass resonances decaying into two partons,
which are observed as a high-mass dijet system. In an analysis performed by the
ZEUS Collaboration~\cite{highmass}, no resonances are observed over the QCD 
background and upper limits on resonance production are presented. 
In this analysis, 
a 95\% C.L. limit on the cross section for $Z^0$ photoproduction at HERA
has been obtained for the first time, 
$\sigma_{e^+p \rightarrow e^+ Z^0 X} < 5.9$\,pb, which is, however, 
well above the cross section of around 0.3\,pb
predicted by the Standard Model.

\subsection{Multijet Photoproduction}

The photoproduction of more than two jets per event is 
related to higher order interactions, but is also sensitive to 
multi-parton interactions (MPI), i.e.
photon-proton
collision in which more than one hard scattering is present at the parton level. 
The study of multijet production is sensitive to MPI, 
which at present cannot be described from
first principles, but have to be modelled 
phenomenologically~\cite{jimmy,mpipythia}.

The ZEUS Collaboration has measured the production of four jets 
in $\gamma p$ interactions in order to study the sensitivity to 
multi-parton interactions~\cite{multipartonzeus}. In this analysis, the
four-jet system is reduced to a three pseudo-jet system by combining the
two jets with the lowest invariant mass in order to exploit the advantages
of the variables used in the three-jet events studied 
previously~\cite{tresjets}.

\begin{figure}[t]
\begin{center}
\mbox{\includegraphics[width=.45\textwidth]{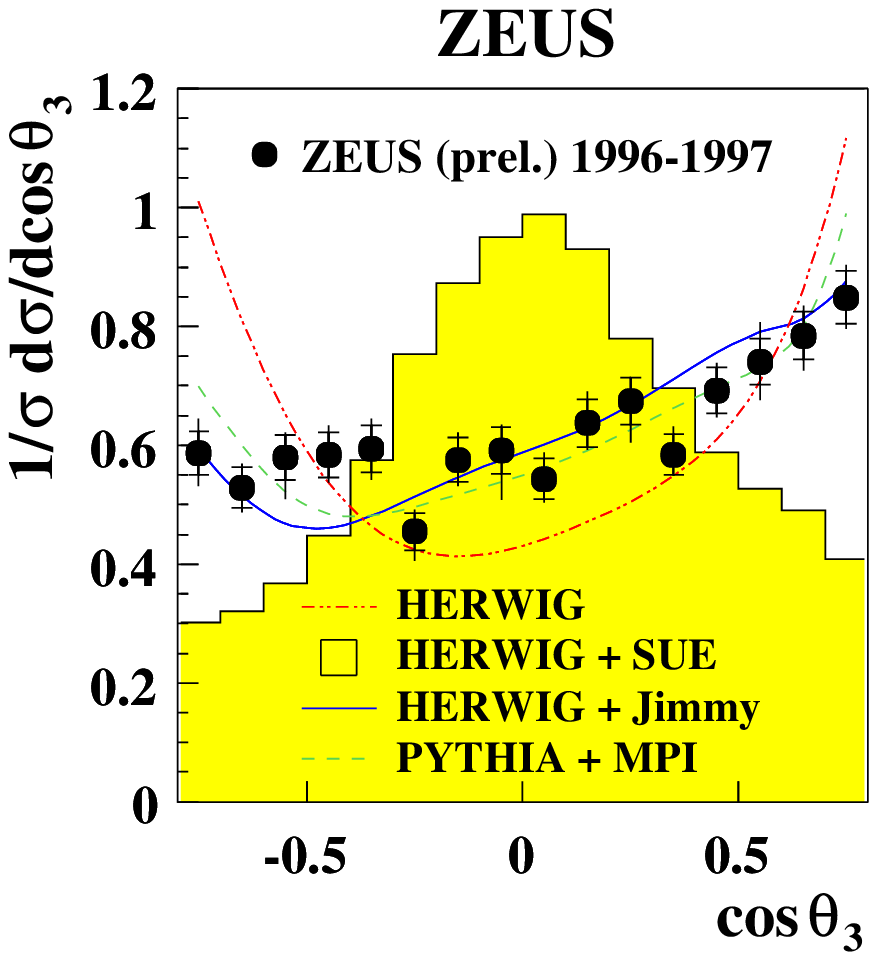} \hspace{0.5cm}
\includegraphics[width=.45\textwidth]{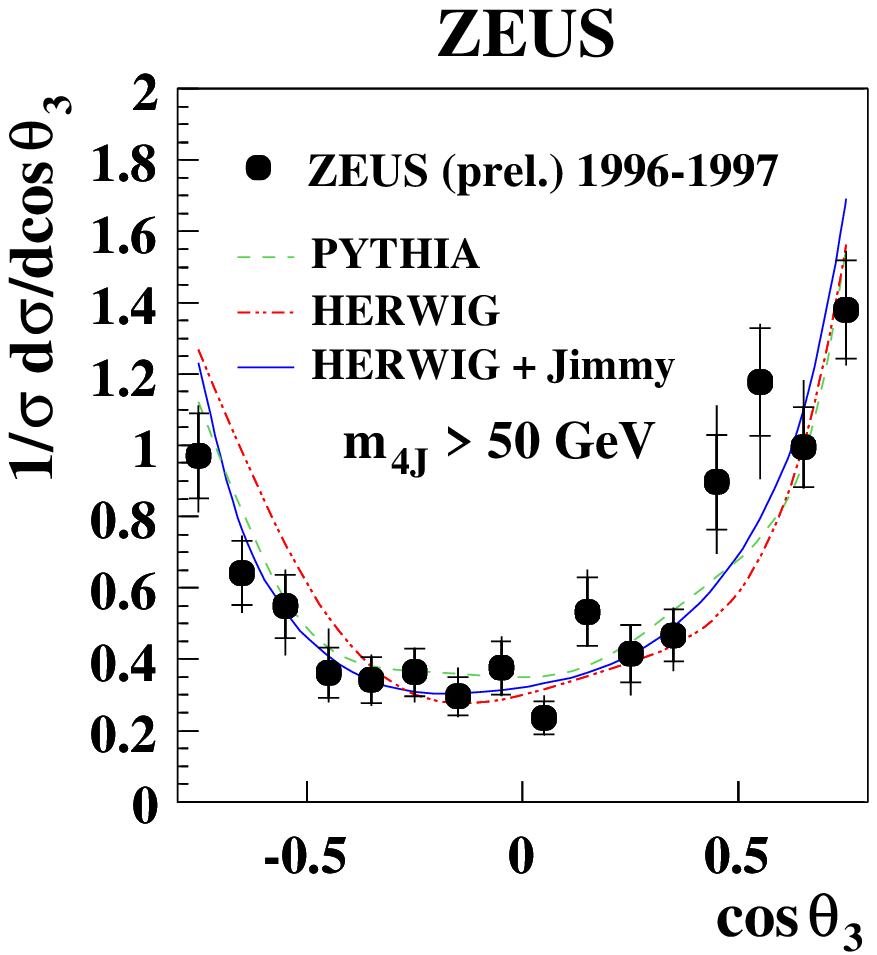}
}
\end{center}
\vspace{-0.2cm}
\caption[]{The $\cos \theta_3$ distribution for the inclusive four-jet sample~(left) 
and for the high-mass, $m_{4J}>50$\,GeV, sample~(right). 
The predictions of HERWIG and PYTHIA with and without including different models 
for MPI are shown}
\label{fig:multiparton}
\end{figure}

The variable used to study the sensitivity to multi-parton interactions 
is $\cos \theta_3$, where $\theta_3$ is the angle between the pseudo jet
with the highest transverse energy and the beam in the four-jet rest frame.
The $\cos \theta_3$ distribution is shown in 
Fig.~\ref{fig:multiparton} for the inclusive sample and for events in which
the invariant mass of the four-jet system, $m_{4J}$, 
is greater than 50\,GeV. 
The inclusive sample
is not described without the inclusion of multi-parton interactions. However,
the precision of the data is not yet high enough to distinguish between the two 
models considered, namely the PYTHIA+MPI~\cite{mpipythia}
and the HERWIG+Jimmy~\cite{jimmy} models.
The soft underlying event in HERWIG is not able to
describe the data in this kinematic region, indicating that the presence
of a soft underlying event cannot account for the results.

For high $m_{4J}$, the sensitivity to multi-parton
interactions becomes smaller and all models
give a reasonable
description of the data.

\subsection{Inclusive Jet Photoproduction}

Inclusive photoproduction of jets allows additional tests 
of QCD to be made
with smaller 
theoretical uncertainties
than in dijet production. On the other hand, the sensitivity
to the partonic content of the photon is smaller and the present 
experimental and theoretical precision does not allow 
discrimination between different parametrisations of the photon pdfs.

The results by the H1 and ZEUS Collaborations~\cite{inclphp}
have been used to test the QCD predictions. NLO QCD calculations 
provide a very good description of the data, 
both in normalisation and shape, over several orders of magnitude. 

\section{Jet Physics in DIS}

The study of 
jet production in DIS is
complementary to jet photoproduction and provides direct access to the underlying
parton dynamics. In this regime, the exchange boson is 
virtual so that the virtuality, $Q^2$, provides a
second hard scale in addition to the transverse energy of the produced 
jets. 

Jet production in DIS is particularly sensitive to QCD effects if the 
jets are produced with high transverse energy in the Breit
frame~\cite{breit}. 
In this frame, the exchanged virtual boson is purely space-like, with
three-momentum $\vec{q}=(0,0,0,-Q)$. 
The production of jets with high energy in the transverse 
direction is only possible if hard 
partons are radiated. Thus, cross sections for producing high transverse 
energy jets in the Breit frame are directly proportional to the strong coupling
constant, $\alpha_s$,
at the lowest order in the perturbative QCD 
expansion. 

Measurements of jet cross sections in DIS have been
compared to the theoretical predictions 
in order to test QCD predictions at HERA in widely different kinematic 
regions and parton configurations. Some of the results are described in the 
following sections.

\subsection{Jet Cross Sections at Low $Q^2$ and at Forward Angles}

The QCD studies performed by means of jet production in neutral current DIS at HERA 
focus on three topics: exploration of the low $Q^2$ region, 
forward- and multi-jet production and precise tests of QCD predictions.

In the first topic, recent results by the H1 Collaboration on 
the inclusive production of jets~\cite{h1incldis}
in the region $5<Q^2<100$\,GeV$^2$ show that improved 
calculations, possibly by including higher order (e.g. NNLO) 
terms in the perturbative QCD expansion, are needed to understand 
low transverse energy jet production in DIS at low $Q^2$; in this kinematic region 
the data are porrly described by NLO QCD and the theoretical uncertainties are large.

\begin{figure}[t]
\begin{center}
\includegraphics[width=.45\textwidth]{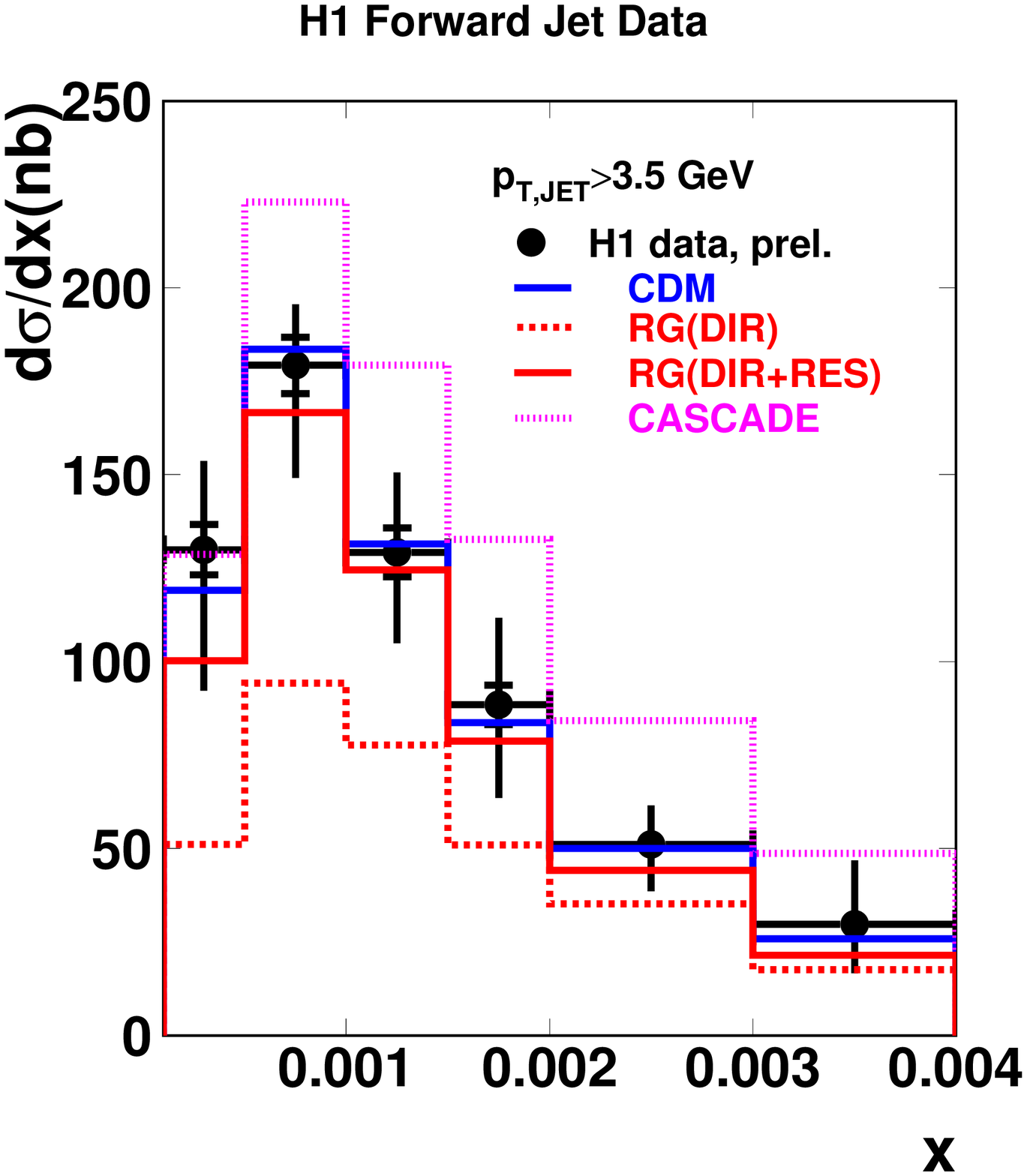}
\hspace{0.5cm}
\includegraphics[width=.45\textwidth]{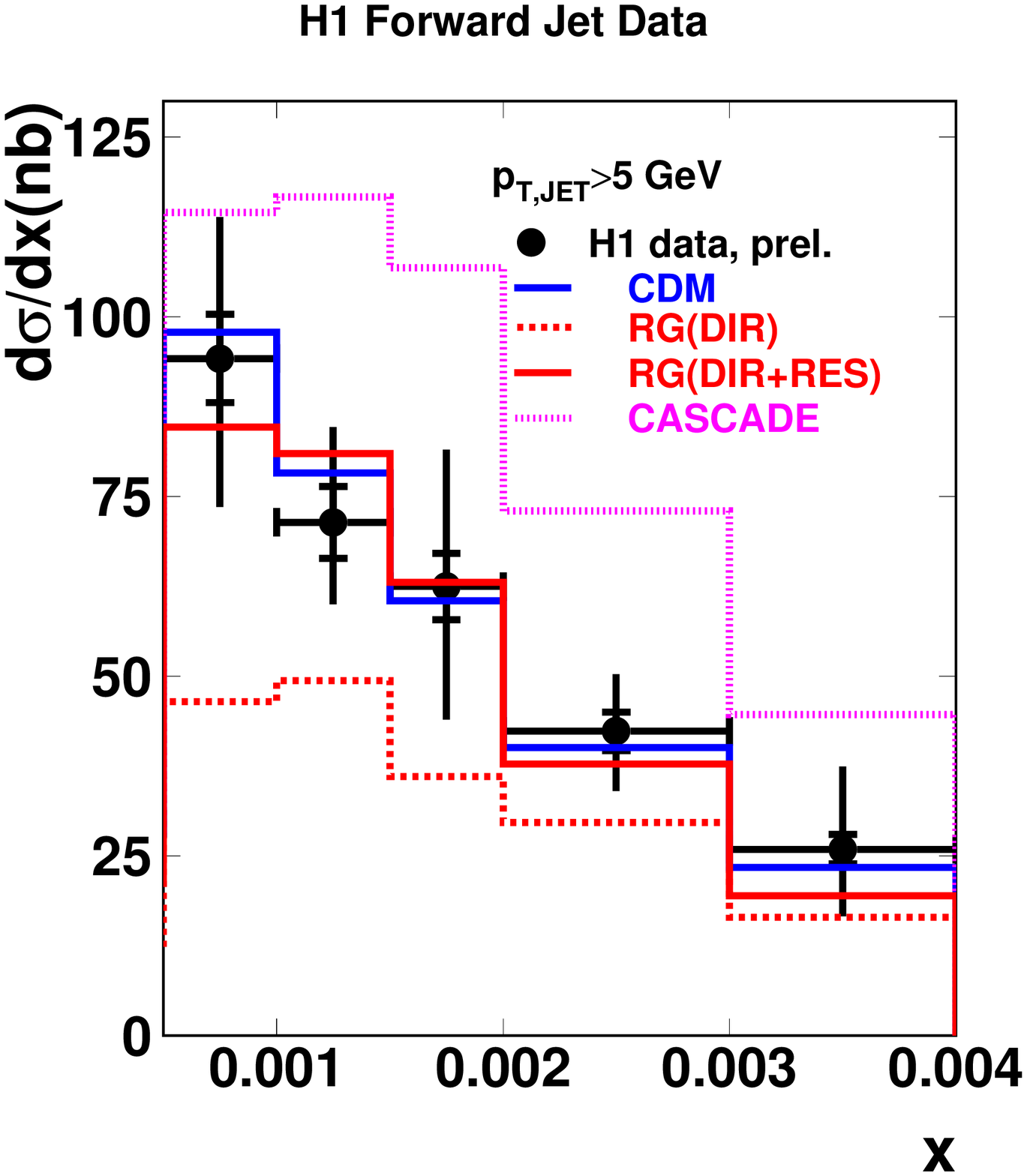}
\end{center}
\vspace{-0.3cm}
\caption[]{Differential cross section 
for forward jet production
as a function of Bjorken $x$ for two regions of jet transverse momentum. 
Also shown are the predictions of several Monte Carlo models}
\label{fig:forjets}
\end{figure}

This disagreement between data and NLO is particularly large 
for jets produced in the forward region, i.e. close to the proton remnant. 
In 
order to study the sensitivity to the evolution scheme, the H1 Collaboration 
has presented the results of a specific analysis~\cite{forjets} on forward jet 
production at HERA. The results are summarised in Fig.~\ref{fig:forjets}, 
where data are presented as a function of Bjorken $x$ and compared to three 
different models. 
The prediction of the colour-dipole model 
as implemented in ARIADNE~\cite{ariadne}, which is close to the BFKL 
evolution scheme, describes the data well, whilst that of RAPGAP~(RG)~\cite{rapgap}, 
using the standard DGLAP evolution, is only able to describe the data when a 
contribution from resolved virtual photons is added.
The CASCADE program~\cite{cascade}, which embodies the CCFM evolution equation,
predicts a rate of forward-going jets which is too high. These conclusions
are independent of the minimum transverse momentum of the selected jets used 
in the analysis.

\subsection{Multijet Production in DIS}

The production of more than two jets with high transverse 
energy in the Breit frame in DIS are related to higher powers of $\alpha_s$ 
in the perturbative QCD expansion. Thus, measurements of multijet 
production in DIS allow the realisation of tests of 
the QCD predictions to higher order in $\alpha_s$. This is the aim of an analysis by 
the H1 Collaboration~\cite{threejets} which shows that the NLO QCD predictions 
provide a good description of the three-jet data over a large range 
in $Q^2$. Furthermore, the ratio of the three- to the two-jet cross section 
is
well reproduced by NLO QCD. The leading-order (LO) 
QCD prediction is not able to describe the measured ratio as a function 
of $Q^2$. 
It should be noted that this type of observable allows the 
performance of precise QCD tests since some experimental and 
theoretical uncertainties cancel out in the ratio.

\subsection{Precise Tests of QCD from Jet Production in DIS}

Recent efforts to understand
both the experimental and the 
theoretical aspect of the analyses of jet production in the Breit frame
in DIS has been
done in order to achieve the high possible precision in the QCD studies
at HERA using jets. The knowledge 
of the jet energy scale at the 1\%--2\% level
has been applied in precise
measurements in kinematic regions in which the perturbative QCD predictions
are well understood.

The most recent analysis~\cite{incljets} was performed by measuring 
inclusive jet production at high $Q^2$
in order to reduce the uncertainties associated to
the low $Q^2$ region and those introduced by the additional
cuts needed in dijet production in order to remove infrared-sensitive
regions~\cite{irsensi}.
The data are well described over many orders of magnitude
by NLO QCD calculations corrected by 
hadronisation effects. 
The differences between data and NLO QCD
are of the 
same size as the theoretical uncertainties. 
This result provides a very
precise test of QCD in describing jet production in DIS.

\begin{figure}[t]
\vspace{-0.5cm}
\begin{minipage}{0.42\textwidth}
\begin{center}
\mbox{\hspace{-.4cm}\includegraphics[width=1.2\textwidth]{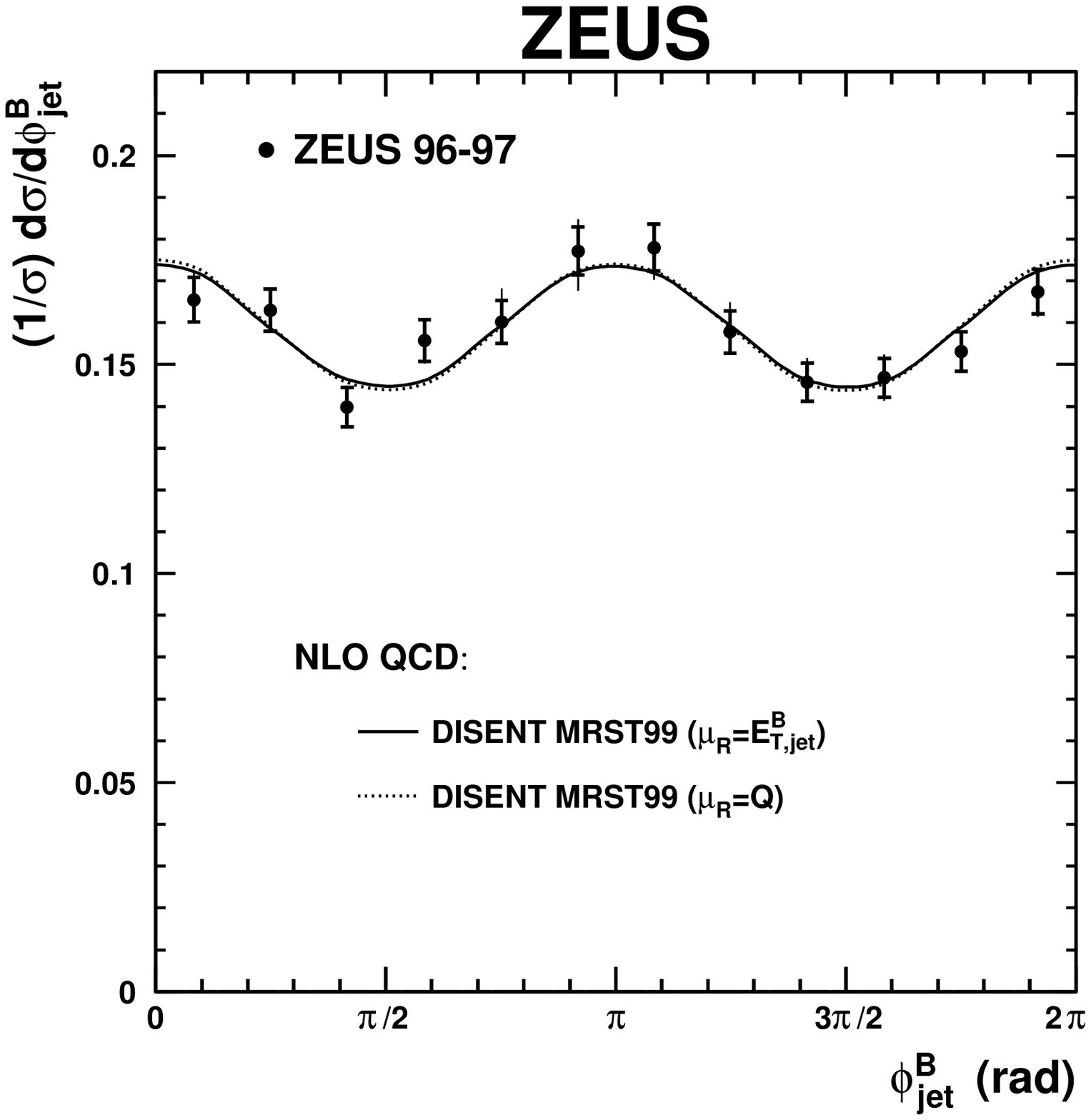}}
\end{center}
\vspace{-0.5cm}
\caption[]{The normalised differential cross section as a function of 
the azimuthal angle in the Breit frame for inclusive jet production.
The NLO QCD predictions are shown for two choices of the renormalisation scale
}
\label{fig:azimut}
\end{minipage}
\hspace{.6cm}
\begin{minipage}{0.5\textwidth}
\vspace{-0.65cm}
\mbox{\hspace{-0.9cm}
\includegraphics[width=1.32\textwidth]{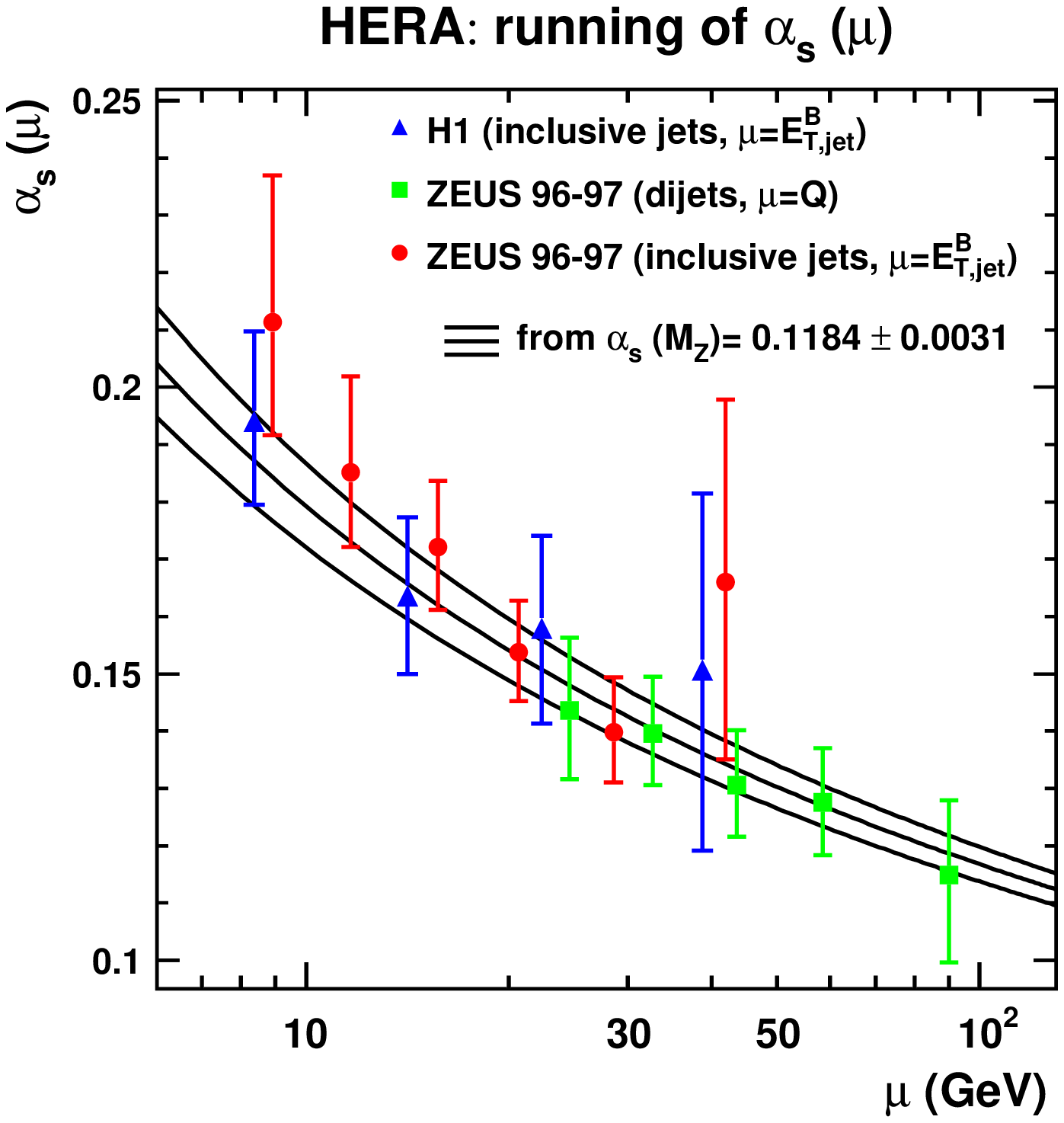}
}
\vspace{-1.0cm}
\caption[]{Values of $\alpha_s$ as measured by H1 and ZEUS at different
energy scales. The values are compared with the running predicted in perturbative 
QCD assuming the world average value for $\alpha_s(M_Z)$}
\label{fig:alphasrun}
\end{minipage}
\end{figure}

Figure~\ref{fig:azimut} shows the distribution of the azimuthal angle of the
jets in the Breit frame at $Q^2>125$\,GeV$^2$, 
performed by the ZEUS Collaboration~\cite{azimut}. 
The measurements are well described by
NLO QCD calculations, which predict a non-uniform distribution 
of the form $A+C\cos 2\phi$~\cite{azimutqcd}. 
This is the first time that the azimuthal asymmetry predicted in QCD has been
observed in DIS using jets.

In addition to the realisation of precise tests of QCD, the
HERA data have been used to determine $\alpha_s$ with a precision comparable 
to the best individual determinations in the world~\cite{incljets,medidasas}. 
The measured values of $\alpha_s$, 
obtained at different scales, presented in 
Fig.~\ref{fig:alphasrun},
display a running which is in good agreement with that predicted
in QCD.

\section{Conclusions}

Jet studies at HERA allow precise tests of QCD predictions which are 
both competitive and complementary to those from other reactions.

With the present precision, jet studies at HERA provide results on 
several topics in QCD, including tests of perturbative 
QCD predictions beyond leading order, studies of the hadronic structure of 
the photon, studies of multijet and forward jet production 
and precise determinations of $\alpha_s$. 

In many of these analyses the uncertainties of the theoretical predictions
are presently limiting the potential of jet physics at HERA.

%

%


\begin{thebibliography}{99.}
\addcontentsline{toc}{section}{References}


\bibitem{h1dijphp} H1 Coll., C.~Adloff et al.: Eur. Phys.J. C \textbf{25}, 13 (2002)

\bibitem{zeusdijphp} ZEUS Coll., S.~Chekanov et al.: Eur. Phys.J. C \textbf{23}, 615 (2002)

\bibitem{highmass} ZEUS Coll., S.~Chekanov et al.: Phys. Lett. B \textbf{531}, 9 (2002)


\bibitem{jimmy} J.M. Butterworth, J.R. Forshaw and M.H. Seymour: Z. Phys. C \textbf{72}, 637 (1996)

\bibitem{mpipythia} T. Sj\"ostrand and M. van Zijl: Phys. Rev. D \textbf{36}, 2019 (1987)

\bibitem{multipartonzeus} ZEUS Coll.: `Multijets in photoproduction at HERA'. Contributed paper to the XXXIst International Conference on High Energy Physics, abstract 849 (Amsterdam, The Netherlands, 2002)

\bibitem{tresjets} ZEUS Coll., J. Breitweg et al.: Phys. Lett. B \textbf{443}, 394 (1998)


\bibitem{inclphp} ZEUS Coll., J. Breitweg et al.: Eur. Phys. J. C \textbf{4}, 591 (1998)\\
                  ZEUS Coll.: `Scaling violations in $\gamma p$ interactions at HERA'. Contributed paper to the XXXIst International Conference on High Energy Physics, abstract 843 (Amsterdam, The Netherlands, 2002)\\
                  H1 Coll.: `Measurement of single inclusive high $E_T$ jet cross-sections in photoproduction at HERA'. Contributed paper to the XXXIst International Conference on High Energy Physics, abstract 1010 (Amsterdam, The Netherlands, 2002)


\bibitem{breit} R.P. Feynman: \emph{Photon-Hadron Interactions} (Benjamin, New York 1972)\\
              K.H. Streng, T.F. Walsh and P.M. Zerwas, Z. Phys. C \textbf{2}, 237 (1979) 

\bibitem{h1incldis} H1 Coll., C.~Adloff et al.: Phys. Lett. B \textbf{542}, 193 (2002)

\bibitem{forjets} H1 Coll.: `Forward Jet production at HERA'. Contributed paper to the XXXIst International Conference on High Energy Physics, abstract 1001 (Amsterdam, The Netherlands, 2002)

\bibitem{ariadne} L. L\"onnblad: Comp. Phys. Comm. \textbf{71}, 15 (1992)

\bibitem{rapgap} H. Jung: \emph{The RAPGAP Monte Carlo for Deep Inelastic Scattering, version 2.08} (Lund University, 2002)

\bibitem{cascade} H. Jung and G.P. Salam: Eur. Phys. J. C \textbf{19}, 351 (2001)\\
                  H. Jung: Comp. Phys. Comm. \textbf{143}, 100 (2001)

\bibitem{threejets} H1 Coll., C.~Adloff et al.: Phys. Lett. B \textbf{515}, 17 (2001)


\bibitem{incljets} ZEUS Coll., S.~Chekanov et al.: Phys. Lett. B \textbf{547}, 164 (2002)

\bibitem{irsensi} M. Klasen and G. Kramer: Phys. Lett. B \textbf{366}, 385 (1996)\\
                  S. Frixione and G. Ridolfi: Nucl. Phys. B \textbf{507}, 315 (1997)\\
                  B. Poetter: Comp. Phys. Comm. \textbf{133}, 105 (2000)

\bibitem{azimut} ZEUS Coll., S.~Chekanov et al.: DESY Report 02-171 (2002). Submitted to Phys. Lett. B

\bibitem{azimutqcd} H. Georgi and H.D. Politzer: Phys. Rev. Lett. \textbf{40}, 3 (1978)\\
                    J. Cleymans: Phys. Rev. D \textbf{18}, 954 (1978)\\
                    G. K\"opp, R. Maciejko and P.M. Zerwas: Nucl. Phys. B \textbf{144}, 123 (1978)\\
                    A. M\'endez: Nucl. Phys. B \textbf{145}, 199 (1978)

\bibitem{medidasas} H1 Coll., C.~Adloff et al.: Eur. Phys. J. C \textbf{19}, 289 (2001)\\
                    ZEUS Coll., J. Breitweg et al.: Phys. Lett. B \textbf{507}, 70 (2001)

\end{thebibliography}
\end{document}